\documentstyle[aps,twocolumn,prb,psfig]{revtex}

\begin{document}

\draft
\title{Potential Energy Surface for H$_2$ Dissociation over Pd(100)}

\author{S. Wilke\cite{prad} and M. Scheffler}
\address{Fritz-Haber-Institut der MPG \\
Faradayweg 4-6, D-14195 Berlin--Dahlem, Germany.}

\date{submitted}

\maketitle

\begin{abstract}
The potential energy surface (PES) of dissociative adsorption of H$_2$ on
Pd(100) is investigated using density functional theory and the full-potential
linear augmented plane wave (FP-LAPW) method.  Several dissociation pathways
are identified which have a vanishing energy barrier. A pronounced dependence
of the potential energy on ``cartwheel'' rotations of the molecular axis is
found. The calculated PES shows no indication of the presence of a precursor
state in front of the surface. Both results indicate that steering effects
determine the observed decrease of the sticking coefficient at low energies of
the H$_{\text{2}}$ molecules. We show that the topology of the PES is related
to the dependence of the covalent H($s$)-Pd($d$) interactions on the
orientation of the H$_{\text{2}}$ molecule.

\end{abstract}

\section{Introduction}

Hydrogen dissociation on metal surfaces has become one of the benchmark
systems for the understanding of dissociative adsorption processes. The H$_2$\
/\ Pd system is particularly interesting as an example of non-activated
dissociation over a transition metal surface. The dissociation process has
been studied using molecular beam adsorption experiments~\cite{ren88} as well
as state resolved time-of-flight measurements of desorbing hydrogen
molecules~\cite{schrot89,schrot91,schrot92a,schrot92b}.

Hydrogen molecules dissociate spontaneously at Pd(100), i.e.  molecules with
low kinetic energy dissociatively adsorb at Pd(100) surfaces with a large
initial sticking coefficient~\cite{ren88,behm80}.  The preference of
non-activated dissociation pathways for low kinetic energies of the H$_2$
molecules is also consistent with the independence of the initial sticking
coefficient on incident angle $\Theta$ as well as with the measured
Maxwell-Boltzmann velocity distribution of desorbing H$_2$ molecules which
corresponds to the surface temperature of the palladium
substrate~\cite{schrot92b,com80}.

There are, however, clear indications that activated pathways exist as
well. After a decrease for beam energies below $\approx$ 0.15\,eV the initial
sticking coefficient continuously increases~\cite{ren88}.  At those larger
kinetic energies the dependence of the initial sticking coefficient on the
incident angle of the molecules changes into a $\cos^{n-1} (\Theta)$ behavior
with $n$ increasing up to $n \approx 1.6$ for beam energies of 0.4~eV.  State
resolved desorption experiments show that the occupancy of the first
vibrational excitation is significantly higher than that corresponding to a
gas of hydrogen molecules in equilibrium with the surface
temperature~\cite{schrot89,schrot92a,schrot92b}. This finding is called
vibrational heating. On the other hand, the rotational degrees of freedom are
occupied less than what the surface temperature would imply (rotational
cooling)~\cite{schrot91}.  In order to explain the vibrational heating, the
values for $n > 1$ in the angular distribution, as well as the increase of the
sticking coefficient with the kinetic energy of the H$_2$ molecules it has
been found necessary in dynamic studies on two-dimensional model
potential-energy surfaces (PES) to assume the presence of a small (about
0.1~eV) barrier in the {\em exit channel}, i.e. in the region of the reaction
pathway where the H-H bond is already significantly
stretched~\cite{schrot92a,dar92}.  Such an energy barrier, however, yields an
additional kinetic energy of thermally desorbing molecules which has not been
measured experimentally~\cite{schrot92a}.

There has been a long debate about the origin of the initial decrease of the
initial sticking coefficient with increasing beam energies. Such a decrease
and the observed dependence of the initial sticking coefficient on the angle
of incidence with a coefficient $n \leq 1$ have been usually assigned to the
presence of a weak, metastable molecular bound state of the H$_2$ molecule in
large distance from the surface (precursor
state)~\cite{ren88,ren94,resch94,prec}. The presence of a precursor state has
also been assumed in order to explain the slow decrease of the sticking
coefficient with hydrogen coverage~\cite{behm80}.  An alternative explanation
relates the initial decrease to the presence of both activated and
non-activated dissociation pathways on the PES~\cite{al90,dix94,gross94a}. At
low kinetic energies molecules with unfavorable orientations may reorientate
during the dissociation process and follow a non-activated pathway but with
increasing kinetic energy this steering is more and more hindered and the
number of molecules proceeding along activated pathways increases.

The fundamental understanding of the dissociative adsorption of hydrogen rests
on the knowledge of the PES which, as will be stressed below, depends rather
cruicially on all six coordinates of the two hydrogen atoms of the H$_2$
molecule. Experimental data have been interpreted in terms of simple, often
two-dimensional model PES as discussed above.  Very recently, accurate density
functional calculation (DFT) of PES~\cite{feib91a,ham94a,whit94,ham95,whit95}
and high-dimensional quantum-dynamical
simulations~\cite{gross94a,gross94,dar94,gross95} of H$_2$ dissociation have
become available thus opening the way for complete {\em ab inito}
investigations of dissociative adsorption and associative desorption
processes. Previous DFT studies of the H$_2$ interaction with metal surfaces
concentrated mainly on system where the dissociation is activated.  There are
a few studies of the interaction of H$_2$ molecules in front of surfaces of
transition metals with a partly filled
$d$-states~\cite{feib91a,ham95,whit95}. In the recent work of White and
Bird~\cite{whit95} investigating hydrogen dissociation on W(100) the presence
of non-activated dissociation pathways was confirmed but in dependence of the
center-of-mass position of the molecule in the surface unit cell even for the
most favorable orientation of the molecular axis activated dissociation has
been obtained as well. Molecules with its axis perpendicular to the surface
have been found to experience no attractive interaction. Summarizing their
results, White and Bird expect molecular steering to play an important role in
the dissociation dynamics at low incident energies.

In this paper we present and discuss detailed calculations of the potential
energy surface for hydrogen dissociation on Pd(100). Here we analyze the
physical properties behind the topology of the PES. We note that the results
are now sufficiently complete for a six-dimensional quantum-dynamical
simulation of the dissociation of H$_2$ molecules on Pd(100)~\cite{gross95}.

\section{Calculation Method}

The calculations have been performed using density-functional theory within
the generalized gradient approximation (GGA)~\cite{per92} and the full
potential linear augmented plane wave method (FP-LAPW)~\cite{wien93,fhiv}.
The application of the GGA is of crucial importance for the calculation of PES
of hydrogen dissociation~\cite{ham94a,fan92,ham93b},but some important
questions concerning the accuracy of PES calculated applying the GGA
remain~\cite{nacht95}.  The good agreement of experimental data with the
results of high-dimensional quantum-dynamical simulation of H$_{\text{2}}$
dissociation based on the calculated PES for H$_2$/Cu(111)~\cite{gross94} and
the PES presented in this paper~\cite{gross95} shows that the GGA reproduces
the important features of these PES with sufficient accuracy.

The FP-LAPW wave functions in the interstitial region are represented using a
plane wave expansion up to $E_{\text{cut}}$=11 Ry, and due to the small radius
of the muffin-tin sphere around H atoms ($r_{\text{MT}}^{\text{H}} = 0.37$
{\AA}) it is necessary to take into account plane waves up to
$\tilde{E}_{\text{cut}}$=169\ Ry for the potential representation~\cite{fhiv}.
Inside the muffin-tin spheres the wave functions are expanded in spherical
harmonics with $l_{\text{max}}=10$, and non-spherical components of the
density and potential are included up to $\tilde{l}_{\text{max}}=3$.  For the
{\bf k}-integration we used 64 uniformly spaced points in the two-dimensional
Brillouin zone corresponding to the c(2$\times$2) surface unit cell.  All
calculations were performed non-relativistically.

In the calculations we used a supercell geometry. The metal substrate is
modeled by three layers slabs which are separated by a 10~{\AA} thick vacuum
region.  Hydrogen molecules are placed at both sides of the slab.  We used a
rigid substrate because due to the large mass ratio between Pd and H the
substrate will not change during the scattering event.  The slab thickness of
only three layers has been chosen with respect to the large computational
demand of the evaluation of the PES. The application of such small layer
thickness for the Pd(100) substrate gives an acceptable accuracy of the PES
because due to the large density of states at the Fermi energy perturbations
in the electronic structure induced by the impinging H$_2$ molecule are
screened very effectively.  We confirmed this by increasing the layer
thickness of the substrate to 5 layers and comparing the total energy with
respect to the separated H$_2$ and Pd(100) surface at various points of the
reaction pathways. We found that in the region of the PES where the H$_2$
molecule approaches the surface and starts to dissociate the differences of
the potential energy are less than 0.05 eV per H$_2$ molecule. In particular,
the character of the dissociation pathway, whether it is activated or
non-activated, is not influenced. This part of the PES determines the dynamics
dissociation of the hydrogen molecule~\cite{gross95}.  The dependence of the
adsorption energy on the layer thickness is more pronounced an the region of
the PES corresponding to adsorbed H atoms and strong H--Pd bonds. Here, the
energy changes with increasing layer thickness up to 0.12 eV per H$_2$
molecule.

The application of such a small layer thickness for the Pd(100) substrate is
possible because due to the large density of states at the Fermi energy
perturbations in the surface electronic structure induced by the impinging
H$_2$ molecule are screened very effectively. We confirmed this by increasing
the layer thickness of the substrate to 5 layers and comparing the total
energy with respect to the separated and Pd(100) surface to that of the three
layer calculations at various points of the reaction pathways. We found that
the differences in the potential energies are below 0.12\ eV and that the
changes are considerably smaller for larger heights of the molecule above the
surface.  For the PES calculation we employed a supercell with a c(2$\times$2)
surface structure.  However, in the case of the geometry where the H$_2$ hits
an on-top site and dissociates into two hollow positions H atoms belonging to
H$_2$ molecules in different surface unit cells within a c(2$\times$2)
structure would come too close during the dissociation process and the unit
cell has been enlarged to a p(2$\times$2) structure.  The convergence of our
results was tested using a p(2$\times$2) surface unit cell, a plane-wave
cut-off for the wavefunctions expansion of $E_{\text{cut}}$=13 Ry and a larger
H muffin tin radius $r_{\text{MT}}^{\text{H}} = 0.48$ {\AA} in the exit
channel region. The resulting changes of the potential energy were again found
to be less than 0.08\ eV per H$_2$ molecule.

The potential energy of a hydrogen molecule as it is used throughout the paper
is defined as the DFT-GGA total energy. The energy zero is the energy of the
geometry difference where the molecule is sufficiently far away from the
surface, ($Z$= 3.7 {\AA}), such that there is practically no interaction
between the molecule and the surface. Zero point corrections and other
vibrations are not included in the PES.

\section{Results}

First, we briefly summarize the properties of bulk Pd, the clean Pd(100)
suface, and adsorbed hydrogen atoms~\cite{wil94,wil95}.  The calculated
lattice constant of Pd calculated by the DFT-GGA is 4.03 {\AA}. This value is
3\ \% larger than the lattice constan obtained within the LDA, and 3.5\ \%
larger than the experimental value~\cite{wil94}.  The clean Pd(100) surface
shows a slight inward relaxation of the topmost Pd layer of $-$0.9 \% of the
bulk interlayer distance. This value is close to the relaxation of $-$0.6 \%
obtained by DFT-LDA calculation~\cite{wil94}.  The equilibrium position of the
hydrogen is the surface hollow site with a small adsorption height $Z_0$=0.1
{\AA}~\cite{wil94}. At the bridge site the adsorption energy is about 0.3\ eV
less favorable and the adsorption height increases to $Z_0$= 1.3 {\AA}.
Adsorption at the on-top position is even unstable against associative
desorption~\cite{wil94}. Obviously, this pronounced dependence of the H -- Pd
bond strength on the geometry of the H adatom effects the potential energy
during the dissociation process as will be discussed below.

\begin{figure}[tb]
\psfig{figure=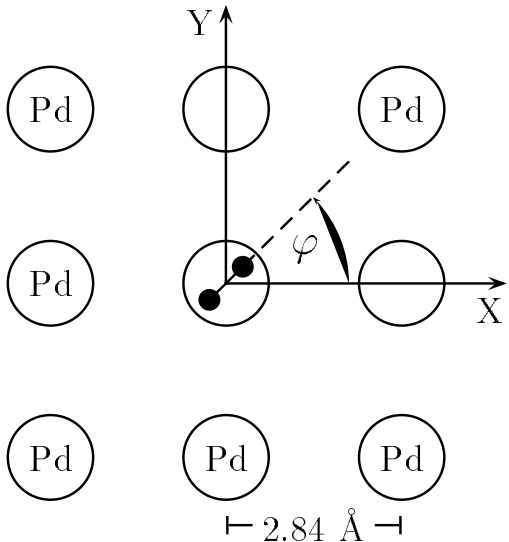 ,width=7cm}
\caption{Coordinate system used to describe the orientation of the H$_2$
molecule with respect to the surface Pd atoms.  The height of the H--H center
of mass above the surface is $Z$, the H--H distance is $d_{\text{H-H}}$, and
the angle of the molecular axis with the surface normal is $\theta$.}
\label{scetch}
\end{figure}

Assuming that the positions of the substrate metal atoms can be kept fixed the
PES of hydrogen dissociation depends on the six coordinates defining the
position of the two hydrogen atoms of the H$_2$ molecule.  Figure~\ref{scetch}
shows the coordinate system used to describe the position and orientation of
the H$_2$ molecule.  The center of mass position is given by the three
cartesian coordinates ($X$, $Y$, $Z$) for which the origin is at the center of
a surface Pd atom.  The two rotational degrees of freedom are given by the
angle of the molecular axis with the $Z$-axis $\theta$ (cartwheel rotation)
and with the $X$--direction $\varphi$ (helicopter rotation). The distance
between the hydrogen atoms is denoted by $d_{\text{H-H}}$.

\begin{figure}[tb]
\psfig{figure=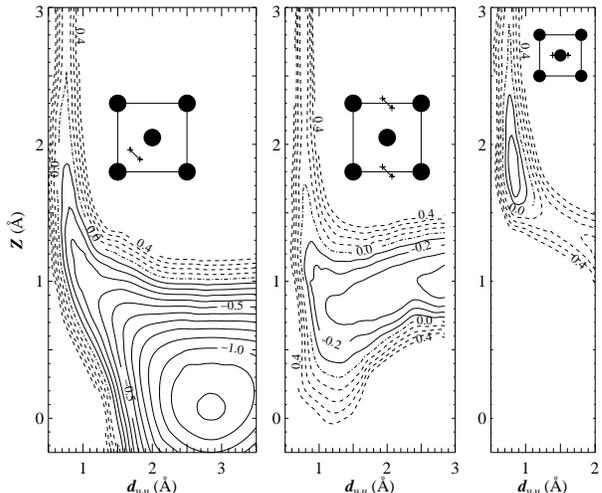 ,width=8cm}
\caption{Cut through the six-dimensional potential energy surface (PES) of a
H$_{2}$ molecule in front of Pd(100). We display ``elbow plots'' where $Z$ is
the height of the H$_{2}$ center of mass over the surface, and
$d_{\text{H-H}}$ is the distance between the two hydrogen atoms. The cut is
defined by keeping the molecule parallel to the surface at an azimuthal
orientation shown in the inset.  The units of the potential energy are eV and
the interval between adjacent contour lines is 0.1 eV.}
\label{pes}
\end{figure}

In order to investigate possible non-activated pathways Fig.~\ref{pes} gives
the PES as functions of H--H distance, $d_{\text{H-H}}$, and the molecular
center-of-mass distance from the surface, $Z$, for fixed, characteristic
positions of the molecules center-of-mass in the surface unit cell: the bridge
($X$=0, $Y$=0.5 $d_{\text{Pd-Pd}}$, $\varphi$=$0^\circ$))(a), the hollow (b)
($X$=$Y$=0.5 $d_{\text{Pd-Pd}}$, $\varphi$=$0^\circ$) and the on-top
($X$=$Y$=0, $\varphi$=$45^\circ$)(c) site. The molecular axis is kept parallel
to the surface ($\theta$=$90^\circ$). For all geometries investigated in
Fig.~\ref{pes} this $\theta$=$90^\circ$ orientation is the energetically most
favorable one (see also Fig.~\ref{rot} below). The dissociation over the
\begin{figure}[tb]
\psfig{figure=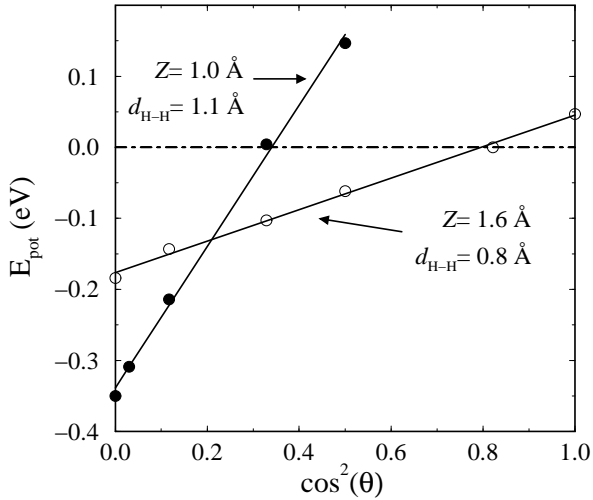 ,width=8cm}
\caption{Dependence of the potential energy of an H$_2$ molecule on the on the
angle $\theta$ of its molecular axis with the surface normal.  The geometry
for $\theta$=90$^\circ$ corresponds to Fig.~2a. The molecule is in the
entrance channel at ($Z$=1.6 {\AA}, $d_{\text{H-H}}$=0.8 {\AA}) (open circles)
or in the exit channel at ($Z$=1.0 {\AA}, $d_{\text{H-H}}$=1.1 {\AA}) (filled
circles)}
\label{rot}
\end{figure}
bridge (Fig~\ref{pes}a) and over the hollow site (Fig.~\ref{pes}b) is
non-activated and proceeds with a continuous gain of energy without a
hampering energy barrier. The dissociation of an hydrogen molecule over the
on-top site with fixed coordinates ($X$=0, $Y$=0) is activated.  The barrier
has an energy of about 0.15 eV and is situated in the exit channel region.
Figure~\ref{pes} shows that non-activated dissociation of H$_2$ molecules on
Pd(100) is possible over several non-activated pathways including a large
region of center-of-mass positions of the hydrogen molecule in the surface
unit cell.

The two-dimensional cuts through the PES in Fig.~\ref{pes} refer to the
situation that the orientation of the molecular axis is {\em fixed}. In
reality the impinging hydrogen molecule rotates.  In the following
we discuss the dependence of the potential energy on the orientation of the
molecular axis, i.e. on the angles $\theta$ and $\varphi$.

The change of the orientations the molecular axis with respect to the surface
normal, $\theta$, and away from $\theta$=90$^{\circ}$, implies a significant
increase of the potential energy.  Figure~\ref{rot} displays the change of the
potential energy of a H$_2$ molecule in dependence of the angle $\theta$ of
the molecular axis with the surface normal for two positions in the PES of
Fig.~\ref{pes}a; all other parameters beeing fixed.  For both distances,
$Z$=1.6 {\AA} and $Z$= 1 {\AA} the potential energy increases like
$\cos^2(\theta)$.  Fig.~\ref{rot} shows that with decreasing distance to the
surface a minimum in the dependence of the potential energy with angle
$\theta$ with increasing depth develops. The ``cartwheel'' rotations are more
and more hindered. There is, however, a large difference between the
``entrance'' channel region, where the hydrogen molecule approaches the
surface and the H$_2$ bond is still intact and the ``exit'' channel region
where the H--H bonds is stretched and the dissociation of the molecule
starts. In the entrance channel, at larger heights $Z$ (see open circles in
Fig.~\ref{rot}), there is no or only a small repulsive interaction of the
molecule with the surface even for a configuration with the molecular axis
perpendicular to the surface $\theta$=$0^\circ$.  Thus, the hydrogen molecules
may approach the Pd(100) surface without an appreciable loss in energy and
independent of their initial rotational state.  In the exit channel where
bonds of the hydrogen atoms with the surface form the potential energy depends
sensitively on the orientation $\theta$ (see filled circles in Fig.~\ref{rot})
and the interaction of the molecule with the surface becomes repulsive already
for small deviations from $\theta$=90$^\circ$.  These conclusion are confirmed
for the extreme case of a hydrogen molecule hitting the surface at the hollow
site with its axis fixed perpendicular to the surface. We find that in the
entrance channel the downmost hydrogen atom may approach the surface plane
without a substantially barrier but for {\em fixed} orientation of the
molecular axis at $\theta$=$0^\circ$ the potential energy increases
drastically if the H--H bond is stretched and a nearly classical hard wall
potential blocks the exit channel.  Without any steering of the molecular axis
during the dissociation non-activated dissociation pathways are possible only
for molecules entering the exit channel region within a narrow region of
orientations $\theta$ of the molecular axis and low initial sticking
coefficient would result.

We expect, however, that the steering of the molecular axis toward
energetically favorable orientations is rather effective. At low kinetic
energies the rotational period of the molecule is comparable to the time the
molecule spends at distances $Z$ corresponding to the end of the entrance
channel. A molecule with a kinetic energy of $E_{\text{kin}}$=0.05\ eV travels
0.6 {\AA} per 0.03 ps and when it is in the $j$=2 rotational state it rotates
during this time (or pathway) by about 90$^\circ$.  The large increase of the
potential energy with angle $\theta$ in Fig.~\ref{rot} at the end of the
entrance channel corresponds to large forces aligning the molecular axis
towards orientations where the dissociation becomes energetically favorable
(``steering effect'').  With increasing beam energy in the adsorption
experiments~\cite{ren88} both the kinetic and rotational energy of the
molecules increases. This shortens the time the molecule spends at the end of
the entrance channel and it becomes more likely that the molecule is reflected
before the axis is aligned. An increase of the rotational energy implies that
states with shorter rotational periods are occupied and those molecules turn
out of the favorable orientation again more easely. As a consequence of the
reduced steering the inital sticking coefficient reduces~\cite{ren88,gross95}.

In the high symmetry configurations shown in Fig.~\ref{pes} the most favorable
orientation of the H$_2$ molecule is that where the molecular axis is parallel
to the surface ($\theta$=$90^\circ$).  For other center-of-mass positions and
orientations $\varphi$ of the molecule the minimum of the potential energy
with respect to the angle $\theta$ deviates from $\theta$=$90^\circ$.  As an
example, if the center of mass of the molecule is between the hollow and
bridge position ($X$=0.25 $d_{\text{Pd-Pd}}$, $Y$=0.5 $d_{\text{Pd-Pd}}$,
$\varphi$=$0^\circ$) the optimal angle at a height $Z$=1.1 {\AA} is
$\theta$=65$^\circ$, i.e. the H atom pointing toward to hollow site is closer
to the surface than that pointing toward the bridge site. This tilting of the
molecular axis is consistent with the different adsorption height of hydrogen
adatoms at the hollow and bridge adsorption site. The potential energy is -0.3
eV, a value between that found along the reaction pathway for a center of
mass-position over the bridge (Fig.~\ref{pes}a) and hollow site
(Fig.~\ref{pes}b) at this height. Generally, the energetically most favorable
orientation of the molecular axis, $\theta$, optimizes the H--surface bond
strength of both H atoms of the H$_2$ molecule.

In the case of ``helicopter'' rotations a pronounced decreases of the
potential energy is found for geometries where H and Pd atoms get closer than
a typical H--Pd bond length (1.7 {\AA}).  These geometrical constraints affect
different positions in the surface unit cell differently. Dissociation over
the bridge site is non-activated in the geometry shown in Fig.~\ref{pes}a
($\varphi$=0$^\circ$). In the case that the hydrogen atoms are oriented
towards the Pd atoms ($\varphi$=90$^\circ$) helicopter rotations are not
hindered only at large distances $Z$. For this configuration a steep increase
of the potential energy is found already at the end of the entrance channel
($Z$=1.2{\AA}, $d_{\text{H-H}}$=0.9 {\AA}).  The H--Pd distance is 1.54 {\AA}
and the potential energy for H atoms orientated towards the Pd atoms is
$E_{\text{pot}}$=$+$0.25\ eV, i.e. 0.57\ eV higher than in geometry with the H
atoms oriented toward the hollow sites (see Fig~\ref{pes}a). In the case that
the center of mass position of the impinging molecule is over the hollow site
helicopter rotations are not hindered. A substantial increase of the potential
energy for the orientation of the molecule with the H atoms oriented toward
the surface Pd atoms ($\varphi$=45$^\circ$) is found only in the exit channel
where the H--H bond is already broken and both H atoms may easily reorientate
separately.  In difference to the ``cartwheel'' rotations the importance of
the steering of the impinging hydrogen molecule into favorable orientations
$\varphi$ of the molecular axis will depend on the position of the molecule in
the surface unit cell.

Metastable, weakly bound molecular states of H$_{\text{2}}$ which may act as
precursor states to the dissociation of the molecule are commonly assumed to
arise due to an interplay the repulsive Pauli repulsion between the closed
shell electronic configuration of H$_{\text{2}}$ and the tails of the metal
surface electronic states and the attractive van der Waals
interaction~\cite{prec}.  In our calculations there is no indication of local
minimum of the potential energy related to such types of precursor states at
large distances in front of the Pd(100) surface.  Although the long-range tail
of the van der Waals interaction related to polarization fluctuations are not
described within the GGA recent calculations of the PES of He atoms in front
of Rh(100) have shown~\cite{max95}, that the GGA is cabable to adequately
reproduce weakly bound states of adsorbates with a closed shell electronic
configuration.  The absence of a considerable repulsive interaction of the
impinging H$_2$ molecule with the surface at larger distances $Z$ even for
unfavorable orientations of the molecular axis and the large decrease of the
potential energy already at distances $Z \approx$2 {\AA} found in
Fig.~\ref{pes} make this mechanism of the formation of molecular bound states
in front of the metal surface rather improbable.

Local minima in a restricted subspace of the high-dimensional PES, may arise
not only at large distances in front of the surface due to the interplay of
the Pauli repulsion and attractive van-der Waals interaction but also at
distances $Z$ where the direct H--metal interaction dominates. They are a
consequence of the pronounced dependence of the H--metal bond on the
orientation of the molecule and the H--surface-metal-atom distances.  For an
example, the dissociation pathway on top of a Pd surface atom
(Fig.~\ref{pes}c) shows a minimum of $E_{\text{pot}}$ in the entrance channel
for {\it fixed} center-of-mass position of the molecule.

\begin{figure}[tb]
\psfig{figure=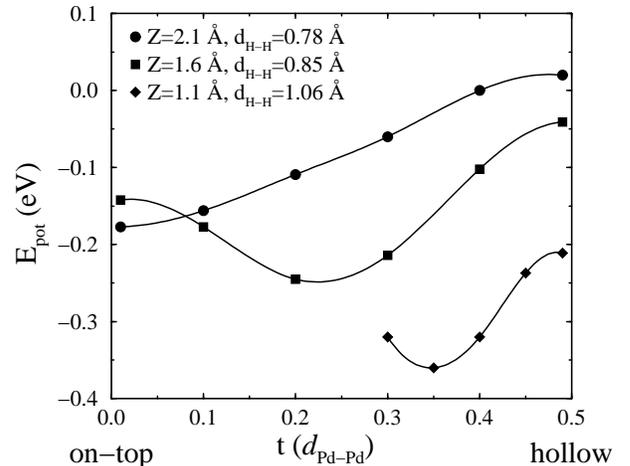 ,width=8cm}
\caption{Dependence of the potential energy of H$_{\text{2}}$ on its
center-of-mass-position along a path $R$=($X$=$-t$, $Y$=$t$, $Z$={\tt
constant}) for three heights of the molecule. The molecular axis is parallel
to the surface ($\theta$=$90^\circ$) and the angle $\varphi$ is fixed at
$\varphi$=$45^\circ$. The point $t$=0 corresponds to the geometry in Fig.~2c.}
\label{xc}
\end{figure}

A more detailed investigation of the dissociation over the on-top site shows
that the minimum of potential energy seen in Fig.~1c is {\em not} a local
minimum in the PES. The molecule can follow a purely attractive pathway if its
center-of-mass position is able to relax in order to optimize the H--Pd
bonding.  Figure~\ref{xc} shows the variation of the potential energy for
different heights $Z$ in the case that the center-of-mass coordinate is
changed along a path, $t$, $R$=($X$=$-t$,$Y$=$t$,$Z$={\tt constant}).  At
large distances $Z > $ 1.8 {\AA} a position on-top of the surface Pd atom
($t$=0) is most favorable. At smaller heights the optimal reaction pathway
involves a motion of the center-of-mass position towards the hollow sites.
This geometry for lower $Z$ is similar to the energetically most favorable
orientation of a H$_2$ molecule in front of a Rh(100) surface~\cite{feib91a}.

It is interesting to compare the PES of hydrogen dissociation on
W(100)~\cite{whit95} with that on Pd(100). Both substrates are transition
metals but tungsten and palladium crystalize into different lattice types (W
is bcc, and Pd is fcc). Furthermore, W is in the middle of the $5d$ and Pd at
the end of the $4d$ transition metal series.  The PES of hydrogen dissociation
on Pd(100) and on W(100)~\cite{whit95}, however, is qualitatively
similar. Both PES exhibit several non-activated dissociation pathways. The
pronounced dependence of the potential energy on the orientation of the
molecule with repect to the surface metal atoms, especially on ``cartwheel''
rotations, is found also on W(100).  There are also characteristic differences
between both PES. This is already documented in the different equilibrium
adsorption site, which is at the bridge position on W(100) in contrast to the
hollow site on Pd(100).  The geometry of the non-activated dissociation
pathways on both surfaces is different. For example the dissociation with
fixed center of mass position over the hollow site has an energy barrier of
0.3 eV on W(100) but no barrier on Pd(100).  These differences will influence
details of the dynamics of the hydrogen molecules on the PES.  The
calculations confirm, however, that the central features of the PES of
hydrogen dissociation, e.g. the presence of non-activated dissociation
pathways, are determined by the presence of a partly filled $d$-band in the
electronic structure of both transition metal substrates.

\section{Breaking of the molecular bond and making of new adatom--surface
bonds}

The large reactivity of transition metals for hydrogen dissociation given by
the presence of dissociation pathways with no or only small energy barriers is
actuated by a partly filled
$d$-band~\cite{feib91a,ham95,har85,hof88,ham95b,coh94a,wil95b}. The mechanism
how the $d$-band may affect the energetics of hydrogen dissociation has been
interpreted in different ways.  Within the Harris-Andersson model~\cite{har85}
the repulsive interaction and the formation of barriers for H$_2$ dissociation
on surfaces of simple and noble metals is, based on the work of Zaremba and
Kohn~\cite{zar76}, related the Pauli repulsion between the closed shell
configuration of the H$_2$ molecule and the tails of the $s,p$-waves of the
surface electronic structure. At transition metal surfaces the Pauli repulsion
is suppressed by a low-energy occupation redistribution of electrons out of
the $s,p$-tails into the large density of holes in the $d$-band just above the
Fermi energy. Thus there is no direct interaction with the metal $d$-states
and the dissociation barriers are governed by the interaction at large
distances to the surface.  The analysis of PES of hydrogen dissociation on
various metal surfaces has shown that the top of the energy barriers occurs in
the {\em exit channel}, i.e. at a point were the H--H interaction is
significantly weakened and and strong H--metal bonds are already beeing
formed~\cite{ham95,ham95b}.  Those results, which are not consistent with the
Harris-Andersson model~\cite{har85}, have been consistently explained within a
reactivity model~\cite{ham95,ham95b,lun79} related to the frontier orbital
concept~\cite{hof88,sail84}.  Superimposed to the interaction of the hydrogen
molecule with the $s,p$-states of the metal, which gives rise to energy
barriers found at simple metals, there is a contribution due to the direct
interaction of H$_2$ with the metal $d$-states.  It is repulsive as long as
both the bonding and the antibonding states of the interaction of the H$_{2}$
$\sigma_g$-orbital with the metal $d$-bands are occupied but becomes
attractive in the case that the antibonding states get depleted. For
transition metals with a high density of $d$-states just at and {\em below}
the Fermi level the states which are antibonding with respect to the adsorbate
surface interaction are pushed above the Fermi level. Thus these antibonding
states remain empty level already in the case of a weak H$_2$--surface
interaction and a small or absent dissociation barrier results.

Feibelman~\cite{feib91a} calculated the dependence of the potential energy of
a hydrogen molecule in front of a Rh(100) surface in dependence of the
orientation of the molecule with respect to the surface metal atoms for fixed
heights, $Z$. He found, that the potential energy variations are governed by
the orientation dependence of the bond strength of the H atoms and the metal
$d$-state. The most favorable orientation of the H$_2$ molecule is not that of
the highest symmetry but that optimizing the H--surface bond strength.

Our results confirm the conclusion that the topology of the PES is determined
by the direct interaction of the hydrogen molecule with the electronic states
of the metal surface. The PES reflects the subtle balance between the
weakening of the H-H bond in the molecule and the energy gain connected with
the rehybridizing into bonds with the metal substrate~\cite{feib91a,lun79}.
Therefore, it is not unexpected that a correlation exists between the energy
at a given height, $Z$, along the reaction path and the energy and optimal
adsorption height of a hydrogen adsorbate at the position of the atoms of the
H$_2$ molecule.  The end of the entrance channel regions in Fig.~\ref{pes}
and, hence, the height $Z$ of a hydrogen molecule at which a significant
stretching of the intermolecular H--H distance occurs, correlates with the
adsorption height of a single hydrogen adatom at the center of mass position
($X$, $Y$)~\cite{wil94}.  At large distances, $Z \, >$ 1.9 {\AA},
center-of-mass positions of the hydrogen molecule on-top of a Pd atom are
energetically most favorable and center-of-mass position over the hollow site
have the lowest gain in potential energy.  The direct dissociation with the
center-of-mass of the molecule over the bridge site into two neighboring
hollow sites ($d_{\text{H-H}}$= 2.85 {\AA}) in Fig.~\ref{pes}a is connected
with a large, gain in potential energy and a small curvature in the ``elbow''
plot which is consistent with the large difference in adsorption height and
energy between bridge and hollow position~\cite{wil94}. In the case that the
molecule impinges at the hollow site (Fig.~\ref{pes}b) the situation is
reversed. The molecule first closely approaches the surface but during the
dissociation process the hydrogen atoms move upward to the energetically less
favorable bridge positions. Consequently, although the dissociation pathway
shown in Fig.~1b is non-activated as well, the energy gain during dissociation
process of the hydrogen molecule is smaller and the curvature in the ``elbow''
plot is larger. The impact position of the molecule over the on-top site of a
surface Pd atom (Fig~\ref{pes}c) is energetically most favorable for larger
$Z$. However, the H--Pd interaction on-top of a metal atom is not strong
enough to break the H-H bond without the formation of an energy barrier which
is in correspondence with the calculated instability of hydrogen adatoms
against associative desorption at on-top sites~\cite{wil94}.

\begin{figure}[tb]
\psfig{figure=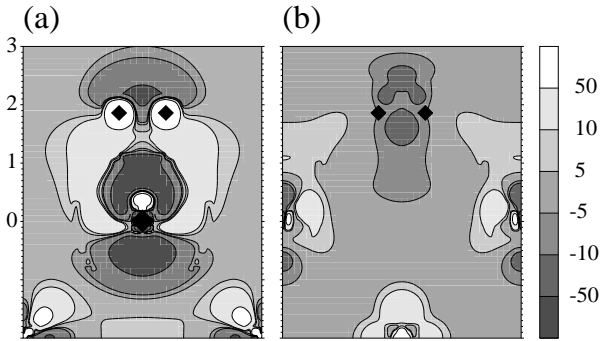 ,width=8cm}
\caption{Difference of the charge density between a hydrogen molecule in front
of Pd(100) at a height $Z$=1.8 {\AA} and that of the superposition of the
charge density of the clean surface and of free hydrogen molecules.  The
center-of-mass position of the molecule is over the on-top site (a) and the
hollow site (b).  The cut plane is perpenticular to the surface and contains
the H$_2$ molecule. The units for the charge density are $10^{-3}$
{\AA}$^{-3}$ and {\AA} for the height above the surface.}
\label{diff}
\end{figure}

A more detailed picture of the interaction of a hydrogen molecule with the
metal surface may be obtained analyzing the electron density induced by a
hydrogen molecule in front of the surface.  In Fig.~\ref{diff} we show the
difference of the electron density between a hydrogen molecule in front of
Pd(100) at a height $Z$=1.8 {\AA} and that of the superposition of the density
of the clean surface and of free hydrogen molecules. Two different
configurations are considered. In Fig.~\ref{diff}a the H$_2$ molecule is over
the on-top site and the potential energy has already a large negative value of
-0.28 eV and corresponds to the minimum in Fig.~\ref{pes}c. In comparison
Fig~\ref{diff}b shows the situation where the H$_2$ molecule is over the
hollow site and the potential energy is close to zero.  Figure~\ref{diff}a
clearly visualizes that the H$_2$--metal surface interaction is connected with
a large polarization of both the H$_2$ molecule and the metal valence
electrons.  The drop in the potential energy in Fig.~\ref{pes}c is caused by a
direct interaction between the $\sigma_{\rm g}$ state of the H$_2$ molecule
and the $3z^2-r^2$-orbital of the surface Pd atom which has a large density of
states just below the Fermi level~\cite{coh94a,wil95b}.  There is an
occupation redistribution {\em inside} the $d$-shell which {\em depletes} the
$3z^2-r^2$-orbital thus reducing the Pauli repulsion. Furthermore, holes are
introduced into the $\sigma_{\rm g}$ state weakening the H-H bond. In addition
an increase of electron density around the center of the H atoms and a more
delocalised accumulation of charge density around the Pd atom occurs
corresponding to an interaction of the antibonding $\sigma_u^{\star}$-orbital
of the hydrogen molecule with metal states just above the Fermi level.  In the
case that the H$_2$ molecule is shifted from the on-top towards the hollow
site in Fig.~\ref{diff}b the interaction with surface Pd $d$-states having a
large density of states just below the Fermi energy is much weaker. The charge
density induced around the hydrogen molecule is now more similar to the
orthogonalization hole found for hydrogen dissociation at the $s,p$-metal
surface Al(110)~\cite{ham93}. But also in this case there is a considerable
depletion of the $3z^2-r^2$-orbitals of the surface Pd atoms and an internal
occupation conversion inside the $d$-shell. Due to the weaker polarization of
the $d$-orbitals the potential energy is increased in comparison to the
position over the on-top site but also in this case no barrier forms.  It is
interesting to note, that the polarization pattern in Fig.~\ref{diff} is close
to that deduced from the electronic properties of the {\em non-interacting}
Pd(100) surface using the reactivity function introduced in
Ref.~\cite{wil95b}.  The induced charge density shown in Fig.~\ref{pes} is
consistent with the reactivity model developed in
Ref.~\onlinecite{ham95,ham95b}.

\section{Conclusion}

We have calculated the PES of hydrogen dissociation on the clean Pd(100)
surface using density functional theory and treating exchange-correlation in
the Generalized Gradient Approximation (GGA)~\cite{per92}. The results confirm
the presence of several, non-activated dissociation pathways. It is shown that
''cartwheel'' rotations of the molecular axis out of the favorable orientation
parallel to the surface are connected with a large increase of the potential
energy, thus confining non-activated dissociation pathways to a small angular
region of molecular axis orientations nearly parallel to the
surface. ``Helicopter'' rotations, on the other hand, are hindered mainly
because of steric restrictions if the H--Pd distance becomes shorter than a
typical H--Pd bonding length. The calculations gave no evidence of a weak
molecular bound state, or precursor state, of the H$_2$ molecule at large
distances in front of the surface.  There is nearly no repulsion of H$_2$
molecules in the entrance channel of the reaction pathways even in the case of
the activated adsorption pathways and thus no appreciable energy is required
for the hydrogen molecule to come close enough to the surface in order to form
hydrogen-metal bonds. The observed pronounced differences in the potential
energy in dependence of the geometry of the molecule with respect to the
surface is governed by the orientation and bond length dependence of the
covalent interaction of the hydrogen atoms mainly with the different Pd
$d$-bands. Summarizing these results it is concluded that the dynamics of the
hydrogen molecule on the high-dimensional PES cannot be understood using
low-dimensional model surfaces but require a quantum dynamic simulation
including all degrees of freedom of the hydrogen molecule. Those calculations
have been performed using the calculated PES presented in this paper as an
input~\cite{gross95}.

\begin{acknowledgements}
Discussions with A. Gross are gratefully acknowledged.
\end{acknowledgements}


\begin{references}

\bibitem[*]{prad} Present Address: Exxon Res. \& Eng. Comp., Annandale, NJ,
08801, USA; e-mail: swilke@erenj.com

\bibitem{ren88} K. D. Rendulic, G. Anger and A. Winkler, Surf. Sci.  {\bf
208}, 404 (1989).

\bibitem{schrot89} L. Schr\"{o}ter, G. Ahlers, H. Zacharias and R.  David,
Phys. Rev. Lett.  {\bf 62}, 571 (1989).

\bibitem{schrot91} L. Schr\"{o}ter, R. David, and H. Zacharias,
Surf. Sci. {\bf 258}, 259 (1991).

\bibitem{schrot92a} L. Schr\"{o}ter, S. K\"{u}chenhoff, R. David, W. Brenig,
and H. Zacharias, Surf. Sci. {\bf 261}, 243 (1992).

\bibitem{schrot92b} L. Schr\"{o}ter, Chr. Tame, R.  David, and H. Zacharias,
Surf. Sci. {\bf 272}, 229 (1992); L. Schr\"{o}ter, Chr. Trame, J. Gauer,
H. Zacharias, R.  David and W. Brenig, Faraday Discuss. {\bf 96}, 55 (1993).

\bibitem{behm80} R. J. Behm, K. Christmann and G. Ertl, Surf. Sci.  {\bf 99},
320 (1980).

\bibitem{com80} G. Comsa, R. David and B. J. Schumacher, Surf. Sci. {\bf 95},
L210 (1980).

\bibitem{dar92} G. R. Darling and S. Holloway, Surf. Sci. {\bf 268}, L305
(1992).

\bibitem{prec} {\em Kinetics of Interface Reactions}, ed. M. Grunze, H.
J. Kreuzer, Springer Series in Surface Science {\bf 8}, (Springer
Verlag,Berlin,Heidelberg,1987), p. 93.


\bibitem{ren94} K. D. Rendulic and A. Winkler, Surf. Sci. {\bf 299/300}, 261
(1994).

\bibitem{resch94} Ch. Resch, H. F. Berger, K. D. Rendulic and E.  Bertel,
Surf. Sci. {\bf 316} L1105 (1994).

\bibitem{al90} P. Alnot, A. Cassuto, and D. A. King, Surf. Sci. {\bf 215}, 29
(1990).

\bibitem{dix94} St. J. Dixon-Warren, A. T. Pasteur, and D. A. King, Surf. Rev.
and Lett. {\bf 1}, 593 (1994).

\bibitem{gross94a} A. Gross, J. Chem. Phys. {\bf 102}, 5045 (1995).

\bibitem{feib91a} P. J. Feibelman, Phys. Rev. Lett. {\bf 67}, 461 (1991).


\bibitem{ham94a}  B. Hammer, M. Scheffler, K. W. Jacobsen and J. K.
N{\o}rskov, Phys. Rev. Lett.  {\bf 73}, 1400 (1994).

\bibitem{whit94} J. A. White, D. M. Bird, M. C. Payne and I. Stich, Phys. Rev.
Lett. {\bf 73}, 1404 (1994).

\bibitem{ham95} B. Hammer and M. Scheffler, Phys. Rev. Lett. {\bf 74}, 3487
(1995).

\bibitem{whit95} J. A. White, D. M. Bird and M. C. Payne, submitted to
Phys. Rev. B.

\bibitem{gross94} A. Gross, B. Hammer, M. Scheffler, and W. Brenig, Phys. Rev.
Lett.{\bf 73}, 3121 (1994).

\bibitem{dar94} G. R. Darling and S. Holloway, J. Chem. Phys. {\bf 101}, 3268
(1994).


\bibitem{gross95} A. Gross, S. Wilke, and M. Scheffler, Phys. Rev. Lett. {\bf
75}, 2718 (1995)


\bibitem{per92} J. P. Perdew, J. A. Chevary, S. H. Vosko, K. A.  Jackson,
M. R. Pederson, D. J. Singh and C. Fiolhais, Phys. Rev. B {\bf 46} , 6671
(1992).

\bibitem{wien93} P. Blaha, K. Schwarz, and R. Augustyn, WIEN93, Technical
University of Vienna 1993.

\bibitem{fhiv} Our force calculation follows the description of R. Yu,
D. Singh, and H. Krakauer, Phys. Rev. B {\bf 43}, 6411 (1991). Further details
are given in B. Kohler, S. Wilke, and M. Scheffler, to be published.

\bibitem{fan92} L. Fan and T. Ziegler, J. Am. Chem. Soc. {\bf 114} (1992)
10890.

\bibitem{ham93b} B. Hammer, K. W. Jacobsen, and J. K. N{\o}rskov,
Phys. Rev. Lett. {\bf 70} (1993) 3971.


\bibitem{nacht95} P. Nachtigall, K. D. Jordan, A. Smith and J. J\'{o}nssson,
J. Chem. Phys., submitted.

\bibitem{wil94} S. Wilke, D. Hennig, and R. L\"{o}ber, Phys. Rev. B {\bf 50},
2548 (1994).

\bibitem{wil95} S. Wilke and M. Scheffler, Surf. Sci. {\bf 329}, L605 (1995).

\bibitem{max95} M. Petersen, S. Wilke, P. Ruggerone, B. Kohler and
M. Scheffler, to be published.

\bibitem{har85} J. Harris and S. Andersson, Phys. Rev. Lett. {\bf 55}, 1583
(1985); J. Harris, Langmuir {\bf 7}, 2528 (1991).


\bibitem{hof88} R. Hoffmann, {\em Solids and Surfaces: A Chemist's View of
Bonding in Extended Structures}, (VCH Verlag, Weinheim, 1988);
Rev. Mod. Phys. {\bf 60}, 601 (1988).

\bibitem{ham95b} B. Hammer and J. K. N{\o}rskov, preprint

\bibitem{coh94a} M. H. Cohen, M. V. Ganduglia-Pirovano and J.  Kudrnovsk\'{y},
Phys. Rev. Lett. {\bf 72}, 3222 (1994).

\bibitem{wil95b} S. Wilke and M. Scheffler, unpublished.


\bibitem{zar76} E. Zaremba and W. Kohn, Phys. Rev. B{\bf 15}, 1769 (1976).


\bibitem{lun79} B. I. Lundqvist, O. Gunnarson, H. Hjelmberg, and
J. K. N{\o}rskov, Surf. Sci. {\bf 89} 196 (1979).


\bibitem{sail84} J.-Y. Saillard and R. Hoffmann, J. Am. Chem. Soc. {\bf 106},
2006 (1984).



\bibitem{ham93} B. Hammer, K. W. Jocobsen, and J. K. N{\o}rskov, Phys.
Rev. Lett. {\bf 70}, 3971 (1993).




\end{references}
\end{document}